\documentclass{PoS}

\title{CP-conserving and CP-violating properties in semileptonic $B_s$ decays with the D\O\ experiment}

\ShortTitle{CP-conserving and CP-violating properties in semileptonic $B_s$ decays with the D\O\ experiment}

\author{\speaker{Lars SONNENSCHEIN}%
        \\
        RWTH Aachen IIIA\\
        E-mail: \email{lars.sonnenschein@cern.ch}}

\abstract{A search for CP violation has been performed in a sample of semileptonic $B_s$ 
          decays corresponding to approximately 5~fb$^{-1}$ of data collected by the D\O\
          detector in Run~II at the Fermilab Tevatron collider. A time-dependent fit to 
          the distributions of $B_s$ candidates yields the flavour-specific asymmetry 
          $a^s_{fs} = [-1.7 \pm 9.1 (\mbox{stat})^{+1.2}_{-2.3} (\mbox{syst})] \times 10^{-3}$, 
          corresponding to the most precise measurement to date for this CP violation 
          parameter. Furthermore a search for the semi-inclusive process $B_s$ to 
          $D_s^{(*)}D_s^{(*)}$ has been performed on a data sample of 2.8~fb$^{-1}$.
          $26.6 \pm 8.4$ signal events are observed with a significance of 3.2 standard 
          deviations above background, leading to a branching ratio of 
          $0.035 \pm 0.010 (\mbox{stat}) \pm 0.011 (\mbox{syst})$. Under certain 
          theoretical assumptions, these double-charm final states saturate CP-even 
          eigenstates in the $B_s$ decays, resulting in a width difference of
          $\Delta\Gamma_s^{CP} / \Gamma_s = 0.072 \pm 0.021 (\mbox{stat}) \pm 0.022 (\mbox{syst})$.
         }

\FullConference{European Physical Society Europhysics Conference on High Energy Physics\\
        July 16-22, 2009\\
        Krakow, Poland}

\begin{document}

\section{Introduction}
The search of large CP violation in the $B_s^0 - \bar{B}_s^0$ system is of special interest
since its observation provides direct indication of new physics. A non-zero CP violation 
weak phase $\phi_s$  arises due to a difference between the $B_s^0 - \bar{B}_s^0$ mixing
amplitude and the amplitudes of the subsequent $B_s^0$ and $\bar{B}_s^0$ decays.
The decay width differences $\Delta\Gamma_s=\Gamma_L-\Gamma_H$ of the light and heavy  
eigenstates and $\Gamma_s^{CP}=\Gamma_s^{\mbox{\scriptsize even}}-\Gamma_s^{\mbox{\scriptsize odd}}$ of the CP eigenstates can be related to the possible presence of new physics 
by the equality $\Delta\Gamma_s = \Delta\Gamma_s^{CP}\cos\phi_s$. On the other hand
the flavor specific asymmetry $a_{fs}^s$ can be related to the CP violation phase by
$a_{fs}^s=\frac{\Delta\Gamma_s}{\Delta m_s}\tan\phi_s$.
The two $B_s^0$ decay analyses presented here are dedicated to set constraints 
on 
CP violation.

\section{Search for CP violation in semileptonic $B_s^0$ decays}
The two $B_s^0\rightarrow \mu^+D_s^- X$ final states
$D_s^-\rightarrow \phi\pi^-$ (with $\phi\rightarrow K^+K^-$) and $D_s^-\rightarrow K^{0*}K-$
(with $K^{0*}\rightarrow K^=\pi^-$) are considered in this analysis \cite{D0a}
which makes use of an integrated luminosity of 5~fb$^{-1}$. 
Initial state flavour is determined from the opposite side and the final state flavour
from the muon charge of the semileptonic $B_s^0$ meson decay.
A likelihood ratio method is applied to increase the significance of the signal candidate
sample. The flavour specific asymmetries $a_{fs}^s$ (signal), $a_{fs}^d$ and 
$a_{fs}^{\mbox{\scriptsize bkg}}$ (background) are determined from an unbinned likelihood fit
in which enter 
variables such as the visible proper decay length, its error, the invariant
$K^+K^-\pi$ mass and flavour tagging parameters.
To obtain unbiased measurements of the flavour specific asymmetries the probability density 
functions entering into the likelihood take properly into account 
all possible detector asymmetries, such as the north south asymmetry of the detector
and the range out asymmetry which describes the difference in acceptance for 
oppositely charged track magnet polarities.
The flavour specific asymmetries are fitted separately for the $\mu^+\phi\pi^-$ and
$\mu^+K^{0*}K^-$ samples. Their combination yields 
$a^s_{fs} = \left( -1.7\pm 9.1 (\mbox{stat})^{+1.2}_{-2.3} (\mbox{syst}) \right) \times 10^{-3}$, which is consistent with the world average values \cite{PDG} of $\Delta\Gamma_s$, $\Delta m_s$ and $\phi_s$.

\section{Evidence for the Decay $B_s^0\rightarrow D_s^{(*)}D_s^{(*)}$ and a Measurement of
          $\Delta\Gamma_s^{CP}/\Gamma_s$}
The $B_s^0\rightarrow D_s^{(*)}D_s^{(*)}$ decays $D_S\rightarrow \phi\pi$ and
$D_S\rightarrow \phi\mu\nu$ (both with $\phi\rightarrow K^+K^-$) are considered in 
this analysis \cite{D0b} which makes use of an integrated luminosity of 2.8~fb$^{-1}$. 
$\gamma$'s and $\pi^0$'s from $D_s^*$ decays are not identified.
The branching fraction is extracted by normalising the decay 
$B_s^0\rightarrow D_s^{(*)}D_s^{(*)}$ to the decay $B_s^0\rightarrow D_s^{(*)}\mu\nu$.
\begin{figure}
\hspace*{18ex} \includegraphics[width=.5\textwidth]{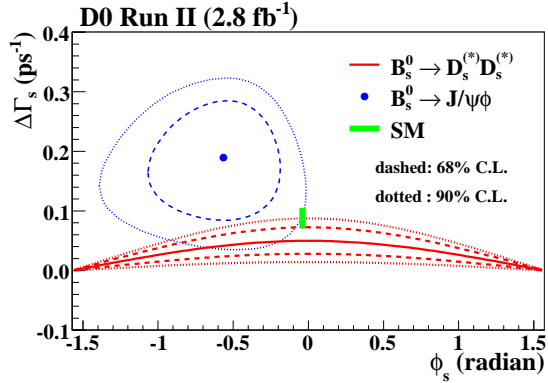}
\vspace*{-2ex}
\caption{ \label{2Dconstraints}
 Constraints in the $\Delta\Gamma_s - \phi_s$ plane.  The solid line represents this
measurement under the theoretical assumptions given in the text. The
dashed and dotted lines correspond to the 68\% and 90\% C.L. intervals respectively.
Contours from the $B_s^0\rightarrow J/\psi\phi$ decay are the equivalent C.L. regions
when measuring simultaneously $\Delta\Gamma_s$ and $\phi_s$. No theoretical uncertainties
are reflected in the plot. 
}
\end{figure}
A two dimensional maximum likelihood fit is applied to the invariant $\phi\pi$ mass 
distribution of the hadronic $D_s$ candidate versus the invariant $KK$ mass of the 
semileptonic $D_s$ candidate. The two dimensional distribution consists of the 
following components:
correlated $D_sD_s$ signal; uncorrelated $D_s\rightarrow\phi\pi$ signal with $D_s$ background,
uncorrelated $D_s$ background with $D_s\rightarrow \phi\mu\nu$, $\phi\rightarrow K^+K^-$ signal 
and correlated $D_sD_s$ background.
The signal template is extracted from a $B_s^0\rightarrow D_s^{(*)}\mu\nu$ data sample.
By extracting the background components the signal yield is estimated to $26.6 \pm 8.4$ events, 
corresponding to a significance of $3.2\sigma$.

The branching ratio $B_s^0$ into two $D_s^{(*)}$ mesons is measured to
${\cal{B}}(B_s^0\rightarrow D_s^{(*)}D_s^{(*)}) =
   0.035 \pm 0.010 \mbox{(stat)} \pm 0.008 \mbox{(exp syst)} \pm 0.007 \mbox{(ext)}$, where
external errors are due to the external input branching ratios taken from the PDG \cite{PDG},
leaving space for further improvements.
Considering the heavy quark hypothesis \cite{HQ} ($\sim 5\%$ theoretical uncertainty) 
along with the Shifman-Voloshin limit \cite{SV} ($3-5\%$ theoretical uncertainty) 
the inclusive final state saturates CP-even eigenstates in the $B_s^0 - \bar{B}_s^0$ system.
This scenario is presented in Fig. \ref{2Dconstraints}, assuming the relation
$\Delta\Gamma_s = \Delta\Gamma_s^{CP}\cos\phi_s$.
Furthermore the mass eigenstates coincide with the CP eigenstates in the Standard Model. 
In this approximation the relative CP decay width difference can be determined to
$\frac{\Delta\Gamma_s^{CP}}{\Gamma_s}\simeq\frac{2{\cal{B}}(B_s^0\rightarrow D_s^{(*)}D_s^{(*)})}{1-{\cal{B}}(B_s^0\rightarrow D_s^{(*)}D_s^{(*)})} = 0.072\pm 0.021 \mbox{(stat)} \pm 0.022 \mbox{(syst)}$. This result is consistent with the Standard Model prediction and the 
world average value \cite{PDG}.
If the CP structure of the final state can be disentangled and the theoretical errors 
can be controlled this measurement can provide a powerful constraint on $B_s^0$ mixing
and CP violation.

\section{Acknowledgments}

Many thanks to the staff at Fermilab and collaborating institutions. 
This work has been supported by the DOE and NSF (USA); CEA and CNRS/IN2P3 (France); 
FASI, Rosatom and RFBR (Russia);
CNPq, FAPERJ, FAPESP and FUNDUNESP (Brazil); DAE and DST (India); Colciencias (Columbia); 
CONACyT (Mexico); KRF and KOSEF (Korea); CONICET and UBACyT (Argentina); FOM (The Netherlands);
STFC (United Kingdom); MSMT and GACR (Czech Republic); CRC Program, CDF, NSERC and WestGrid Project 
(Canada); BMBF, DFG and the Alexander von Humboldt Foundation (Germany); SFI (Ireland); 
The Swedish Research Council (Sweden); and CAS and CNSF (China).

\end{document}